\documentstyle[prl,aps,twocolumn,floats,psfig]{revtex}
\begin{document}
\bibliographystyle{simpl1}
\title{ Guided Simulated Annealing Method for Optimization Problems}
\author{C.~I. Chou$^{1}$, R.~S. Han$^{1}$, S.~P. Li$^{1}$ and T.~K. Lee$^{1,2}$}
\address{$^1$Institute of Physics, Academia Sinica, Taipei, Taiwan 115\\
$^2$ Physics Division, National Center for Theoretical Sciences,
P.O.Box 2-131,Hsinchu, Taiwan 300\\
{\em (Draft on {\today} )}
\medskip ~~\\ \parbox{14cm}{\rm
Incorporating the concept of order parameter of the mean-field
theory into the simulated annealing method, we presented a new
optimization algorithm, the guided simulated annealing method.
  In this method mean-field order parameters are
calculated to guide the configuration search for the global
minimum. Allowing fluctuations and improvement of mean-field
values iteratively, this method successfully identified global
minima for several difficult optimization problems. Application of
this method to the HP lattice-protein model has found a new lowest
energy state for an $N = 100$ sequence that was not found by other
methods before. Results for spin glass models are also presented.
}} \maketitle

Optimization problems (OP's)\cite{hor} arise in areas of science,
engineering and other fields.  As emphasized by the Levinthal's
paradox\cite{lev}, the main difficulty in these OP's is the
exponential increase in the search space with system size.  The
travelling salesman problem\cite{rei}, the protein folding
problem\cite{cre} and the Lennard-Jones microcluster
problem\cite{wal} are examples belonging to this class.

Two of the most popular numerical approaches to treat OP's are the
simulated annealing (SA) method\cite{kir} and genetic
algorithms\cite{gen}. For these approaches to succeed, the methods
must be able to sample as much configuration space (CS) as
possible.  However, most of these OP's have energy landscapes
filled with local minima surrounded by high barriers. Therefore
many sophisticated methods were invented to avoid entrapment in
local minima and to increase efficiency in configuration sampling.
The local landscape paving\cite{han}, basin hopping\cite{wal},
stochastic tunneling\cite{wen}, the various generalized ensemble
methods\cite{ber,oka}, or the nonextensive statistics\cite{tsa}
are all based on this philosophy. Even if the entrapment problem
is resolved, it would not be particularly efficient, to sample the
multi-dimensional CS by radon trials. It would be preferable to
have some guidance about the most probable region where the global
minimum is located.  We will show below that the concepts used in
the familiar mean-field (MF) approach in many-body problems can be
useful in this regard.

In the MF approach, specific physical quantities are identified as
order parameters (ORP's). These ORP's (or just one parameter) are
problem specific and usually carry the most important information
about the system's ground state. They acquire different values
between high temperature states and the ground state.  The idea of
the MF approach is to use the ORP's to lead the many-body system
to low energy states so that injecting small fluctuations would
bring the system to its ground state. Some understanding of the
ground state is necessary to choose the correct ORP's.
Fortunately, we do have information about the results we are
looking for in most cases.  For example, we know that
Lennard-Jones clusters\cite{wal} will have most of its inner core
atoms arranged with certain  symmetry. The density of atoms is
therefore a good ORP.

We incorporate this MF concept into the SA method. In our
approach, the ORP's  with some assumed initial values is used to
bias the search of CS to favor the regions dictated by their
values.  In other words, the ORP is used as a guiding function
(GF) in the search of the ground state. By changing its values
iteratively, the ORP continuously adjusts its values until the
ground state is reached. This combination of SA method with a GF
will be called the guided simulated annealing (GSA) method. As an
illustration of the method, we will apply it to the HP
lattice-protein model\cite{oka,cha} and the spin glass (SG)
model\cite{edw}.
  For the former, we have found a new lowest energy state that
has not been found before for a sequence of length 100. For the
latter new results, for 5-dimensional SG models are found. The
method has also been successfully applied to the x-ray
crystallographic problem for large molecules\cite{ch1} and the
Lennard-Jones cluster problem\cite{ch2}.

Lattice protein models are the simplest models which have been
playing important roles in the theoretical studies of protein
folding.  In these models, protein chains are heteropolymers which
live on two or three dimensional regular lattices.  They are
self-avoiding chains with attractive or repulsive interactions
between neighboring unbonded monomers. In most simulations, people
consider only two types of monomers---the hydrophobic (H) and
polar (P) monomers. The reader is referred to\cite{cha} and
references therein for more detailed discussion. Despite its
simplicity, the number of conformations of a lattice model protein
chain becomes enormous as the length of the chain grows.  It is a
challenging task to find the global minimum and is an ideal test
for the GSA method.

Our approach begins with a population of $M$ randomly generated
conformations. We let each of them evolve independently as in the
usual SA method\cite{kir}. We here adopt the three commonly used
MC moves---the end move, the corner move and the crankshaft
move\cite{oka}.  Aside from these moves, we also include one more
type of move: a rotation of a portion of the chain about a chosen
point of this chain\cite{ung}.  We adopt the Metropolis
rule\cite{kir} for all our MC moves.  In each of the $M$
independent runs, keep a record of its lowest-energy solutions
after a preset number of MC steps.


















Our next step is to construct a GF.  As mentioned above, the
choice of the GF is determined by the ORP which represents an
important property  that is different between low energy and high
energy states. For a real protein,  Ramachandran torsion angles
are clearly  good candidates for the ORP. We therefore consider
the angular distribution at each monomer or more appropriately the
local
 substructure is used as our GF.
Let us consider a chain consists of ten monomers on a two
dimensional square lattice with a conformation as shown in
Fig.1(a). For this protein chain, we take for each time a segment
of 5 monomers.  A total of 6 segments can be identified if we move
along the chain from the first monomer. For each segment, we
record its structure as follows.  As we move along the chain,
there are three different cases we will encounter: go straight
(0), turn left (1) or turn right (-1).  For example, for the first
segment from monomer 1 to 5,
 we go left (1), right (-1), then  straight (0).
 This substructure associated with monomer 3
is denoted as (1,-1,0). The fourth segment from monomer 4 to 8
then has the substructure (-1,1,-1) associated with monomer 6.
According to this classification, there are a total of 25 possible
substructures for a segment of length 5 in two dimensions.  Notice
that (1,1,1) and (-1,-1,-1) both form closed squares and thus are
not allowed.






%












All 25 types of substructures discussed above are for segments
with 5 monomors.  There are two segments of 6 monomors, shown in
Fig.1(b), which deserve special attention. Not only they are
related to the crankshaft MC move, they seem to have special
weight in the structures. Thus, if we have segments with length 6
of one of these two types ( (1,-1,-1,1) or (-1,1,1,-1) ), we will
record it using these two types of substructures instead of the 25
types above. Again, these two additional types of substructures
will be associated with the third monomer of their corresponding
segment\cite{note1}.
 We add up the number of times each substructure
appears at every monomer of the chain for the $M$ lowest-energy
solutions and then make up a set of distribution functions. These
distribution functions, denoted by $p(i,j)$ for $j$th type
substructure at the $i$th monomer, are our GF's for the next layer
of simulation and our values of ORP's in this layer of simulation.
This completes our first layer of MC simulation. In our
discussion, a layer of simulation means a set of $M$ individual SA
runs for a preset MC steps plus the construction of the GF
aforementioned.

There are now a total of 27 types of substructures, hence $\sum_j
p(i,j) =1$, where $ 1 \le j \le 27$. Without interaction between
monomers, $p(i,j)$ is independent of monomer position $i$ along
the chain and $p(i,j)=p_0(j)$\cite{note2}.




In the second layer of simulation, a set of $M$ independent SA
runs is again performed. The GF will now be incorporated in our
search. Unlike usual SA or MC rules, where every monomer has equal
probability to be selected to change its substructure before the
Metropolis rule is applied, we give a higher probability to pick
the substructures within the protein chain which appear less
frequent in the GF's and try to change them into substructures
with a higher probability of appearance in the same set of GF's.
This is very similar to what the ORP's
 do in the MF approach of statistical models. For example,
the assumed MF magnetization in a spin model will greatly bias the
direction of the spins. It should be noted that in order to allow
enough fluctuations, only slightly larger weight should be given
 to the GF.  The values of ORP's or GF
will be modified when a new set of $M$ lowest-energy states is
obtained at the end of the second layer of simulation. To  avoid
using solutions that could already have been locally trapped,  we
always start the new layer simulation
  with $M$ randomly generated conformations.

\begin{figure}
\centerline{\psfig{figure=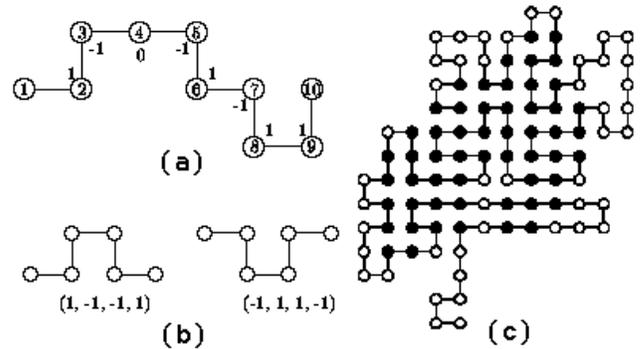,width=3.3in}}
\vspace*{5mm} \caption{(a)A $10$-monomer protein chain on a two
dimensional square lattice. (b)Two types of substructures of
segments with length 6. (c)Conformation of the sequence
$100_{(1)}$ with lowest energy $E=-48$. Solid dot represents the H
monomer. There is only attractive interaction between H and H
nonbonded monomers.} \label{fig4}
\end{figure}

There are numerous ways to apply the GF in our MC process.  We
will here discuss a particular way chosen by us. The first step is
to decide whether to take the rotational move or the three local
MC moves.  A small probability (about $30\%$) is assigned for the
rotational move and the GF is only used for the three local MC
moves.  Before we use the local moves to change the position of a
particular monomer in a certain conformation state, we first
determine the monomers within the chain that are allowed to change
positions. For each of these monomers, we look for the
corresponding substructure of its segment and their GF values,
namely $p(l,j)$.  The probability to select which monomer to be
moved is determined by this GF value. The smaller this value is,
the more likely it should be changed.  Hence we define the
probability to be selected proportional to  $\alpha^{-p_a(l)}$,
where the adjusted distribution function $p_a(l)=p(l,j)-p_0(j)$.
The background distribution $p_0$ for the non-interacting chain is
subtracted from the GF value to signify the contribution by the
interaction and the sequence effect.  The parameter $\alpha>1$ is
determined by tuning the efficiency of the algorithm. For end
monomers we set $p_a=0$.

One then adds up and normalizes all the possible probabilities.
The probability $P(l)$ for monomer $l$ to be considered for a move
is then
$$
\displaystyle P(l) = \frac{\alpha^{-p_a(l)}}{\sum_m
\alpha^{-p_a(m)}}  , \eqno(1)
$$
where $m$ is summed over all allowed monomers. Clearly if
$\alpha=1$, the GF is not used at all. If $\alpha$ is chosen to be
much larger than 1, all the segments are forced to be equal to the
largest GF values and no fluctuations are allowed.  A better
choice is to have  $\alpha$ slightly larger than 1 so that the GF
could be modified during successive layers of iteration. For the
data reported below, we find that  it is best  to have $\alpha$
between 1.2 and 1.3.

Table I lists our GSA result on 2D HP lattice model for sequences
studied by many other groups.  We used a population of $2N$
independent samples in each case, where $N$ is the length of the
chain.  For each independent sample, we started at a certain
temperature and ran for a preset number of MC steps.  We then
lowered the temperature and ran for the same preset number of MC
steps.  A set of 20 different temperatures was used.  For the set
of parameters we used here, a typical run on a sequence with $N =
36$ takes 46 seconds on a Pentium IV 1.4GHz CPU and takes less
than 10 hours for the case with $N = 100$ shown in the table.

\vskip 0.5cm {\footnotesize{\noindent Table I. Results of our
algorithm on sequences of the 2D HP lattice model. \lq\lq
sequence" refers to the length of the sequence.  \lq\lq Layers" is
the number of layers in the simulation.  \lq\lq Steps" is the
number of MC steps at each chosen temperature in each layer.
\lq\lq Previous" is the previous lowest energy states using other
algorithms and \lq\lq Ours" is the lowest energy states obtained
by using our method. }}

\begin{center}
\begin{tabular}{c  c  c  c  c} \hline \hline
Sequence \,\,\,\,\,  & \,\,\,\,\, Layers\,\,\,\,\,  & \,\,\,\,\,
Steps\,\,\,\,\, & \,\,\,\,\,Previous \,\,\,\,\, & \,\,\,\,\, Ours
\\ \hline
36\cite{ung}\,\,\,\,\, & \,\,\,\,\, 2\,\,\,\,\, & \,\,\,\,\, 100\,\,\,\,\, & \,\,\,\,\, -14\cite{ung}\,\,\,\,\, & \,\,\,\,\, -14 \\
48\cite{ung}\,\,\,\,\, & \,\,\,\,\, 2\,\,\,\,\, & \,\,\,\,\, 100\,\,\,\,\, & \,\,\,\,\, -23\cite{tom}\,\,\,\,\, & \,\,\,\,\, -23 \\
60\cite{ung}\,\,\,\,\, & \,\,\,\,\, 2\,\,\,\,\, & \,\,\,\,\, 200\,\,\,\,\, & \,\,\,\,\, -36\cite{fra}\,\,\,\,\, & \,\,\,\,\, -36 \\
64\cite{ung}\,\,\,\,\, & \,\,\,\,\, 4\,\,\,\,\, & \,\,\,\,\, 640\,\,\,\,\, & \,\,\,\,\, -42\cite{beu}\,\,\,\,\, & \,\,\,\,\, -42 \\
85\cite{lia}\,\,\,\,\, & \,\,\,\,\, 2\,\,\,\,\, & \,\,\,\,\, 1700\,\,\,\,\, & \,\,\,\,\, -52\cite{lia}\,\,\,\, & \,\,\,\,\, -52 \\
$100_{(1)}$\cite{ram}\,\,\,\,\, &\,\,\,\,\, 10\,\,\,\, & \,\,\,\,\, 2000\,\,\,\, & \,\,\,\,\, -47\cite{fra}\,\,\,\, & \,\,\,\,\, -48 \\
$100_{(2)}$\cite{ram}\,\,\,\,\, & \,\,\,\,\, 10\,\,\,\, &
\,\,\,\,\,2000\,\,\,\, & \,\,\,\,\, -50\cite{chi}\,\,\,\, & \,\,\,\,\, -50 \\
\hline\hline

\end{tabular}


\end{center}


As indicated in Table I, we have been able to obtain all the
previous best results of the 2D HP chains.  We further obtain the
lowest  energy for a conformation of the sequence $100_{(1)}$ that
was not found by other methods. Its conformation is shown in
Fig.1(c). This is the only conformation we found with this lowest
energy while more than 40 different conformations with energy
$E=-47$ were found. We have also found many conformations of the
sequence $100_{(2)}$ with an energy $E = -50$ that are different
from the one given in\cite{chi}, which can be provided to the
reader upon request.

As shown in Table I, for most sequences, a few layers of iteration
is enough to find the lowest energy except $N = 100$. It should be
noted that the number of layers and MC steps are not tuned to the
optimal speed. In fact, conformations with $E=-50$ for the
sequence $100_{(2)}$ and $E=-47$ for the sequence $100_{(1)}$ are
found within the first seven layers already.  We kept the program
running to find possible lower energy states.  Since the program
is very efficient, we can afford this extra searching.

To understand our results better, we have carefully examined the
topology of local structures in the MC simulations. We found that
local substructures form in the early stage of the search process.
In addition, there is a very strong correlation between types of
segments of sequence with types of substructures. For example, the
$HPH$ sequence segment has an unusual large probability to turn
left or right at $P$ monomers. $HPPH$ sequence almost always has
both $P$ turn left or right together with both $P$ likely  on the
surface. Their substructures are mostly related to structures
shown in Fig. 1(b). Special consideration of these substructures
of length 6 instead of just length 5 is important in identifying
them.

To examine this further, we folded $50,000$ sequences of $36$-mers
chains to their minimal states.  For a total number of $236352$
$HPH$ segments, $93.8\%$ turn left or right at the middle $P$
monomer, which is much larger than the average possibility
$66.7\%$.  Furthermore, most $P$ monomers ($89\%$) are found on
the surfaces of the folded conformations.  For the $HPPH$ segment
(a total of 101155), $92.6\%$ simultaneously turn left or right at
the two $P$ monomers compared to the average possibility of only
$22.2\%$. The possibility that the two $P$s both stay on the
surface is $92.5\%$.  Since these minimal states are not
necessarily  the native states, we also carried out a complete
search for  all the native states for  sequences with $11$ to $17$
HP monomers. Similar results are obtained. For a total of 7373
$HPH$ segments, $99.5\%$ turn left or right at the $P$ monomer,
and $98.8\%$ of the $P$ monomers are on the surface.  For 4289
$HPPH$ segments, $99.7\%$ turn left or right at the same time at
the two $P$s. Only one $P$ monomer is found in the core of the
native structure.
















%

The strong correlations observed above between certain type of
sequence segment and a particular substructure may be responsible
to help us locate the \lq\lq native" state much faster in our
approach. It should be noted that this is consistent with the
recent observation by Baker\cite{Baker} that simple topologies
with mostly local interactions are more rapidly formed than those
with non-local interactions. The GF or the ORP  we used seem to
have captured the importance of local substructures in the protein
structure prediction problem.










%

%











Another example is the SG problem in statistical physics. In
\cite{spl}, we have performed simulations in 3D SG model\cite{edw}
using our GSA method.  We here performed further simulations for
this model in 4D and 5D and present the results in Table II.
During the simulation, the average spin configuration at each site
is kept which is equivalent to the local magnetization, and is
used as our ORP or GF for subsequent layers of simulation. The
reader is referred to \cite{spl} for more detail on how to use our
algorithm in SG models.

\vskip 0.5cm {\footnotesize{\noindent Table II. Tests on 4D and 5D
SG. L is the lattice size.  m is the number of spin configuration
cases and E is the average ground state energy of the m cases. All
simulations are done at $T=1.15$.}}

\begin{center}
\begin{tabular}{c | c | c | c | c | c} \hline \hline
&\multicolumn{3}{c|} { $D=4$}  & \multicolumn{2}{|c} {$D=5$}
\\ \hline
L & m & E & Ref.\cite{boe} & m & E \\ \hline
$3$ & $8000$ &$-2.0249(7)$ & -2.0214(6) & $5000$ & $-2.3168(5)$ \\
\hline
$4$ & $2000$ &$-2.0699(6)$ & -2.0701(4) & $1000$ & $-2.3506(4)$ \\
\hline
$5$ & $1000$ & $-2.0849(5)$ & -2.0836(3) & $50$ & $-2.3530(10)$ \\
\hline
$6$ & $200$ & $-2.0887(7)$ & -2.0886(6) & &  \\
\hline $7$ & $50$ &  $-2.0904(12)$ & -2.0909(12) & &   \\
\hline\hline
\end{tabular}
\end{center}

In our simulation, the number of Monte Carlo steps and layers used
ranges from 300 and 2 (for L=3) to 1400 and 3 (for L=7) in the 4D
case.  About the same set of parameters are used in the 5D case.
The CPU time for a trial run on a SUN 450MHz processor for the 4D
L=7 case is about 400 seconds.  A few trial runs are performed and
the best solution is chosen for each spin configuration.  One can
see that our result is comparable with that of \cite{boe} in the
4D case and has considerable improvement in the 5D case\cite{wan},
where $E_\infty = -2.347(16)$.

In summary, we have presented a new approach to treat general OP's
with continuous or discrete variables.  Based on the idea of MF
theory, the GSA method introduces ORP's.  These ORP's are then
used as a GF to help direct the search of global minimum in the MC
process.  The method is illustrated by applying to
 the HP lattice protein model.
We have found all the putative ground state energies reported for
the chains that we tested.  A new ground state
 for a particular sequence of length 100 has been found.
In addition, strong correlations between particular sequence
segments and substructures are found. We have also discussed
briefly the method and its results in
 the SG  problem.

This GSA method has several special features. It emphasizes biased
search in CS for the global minimum instead of the non-biased
search algorithm used by most other approaches\cite{hor}. This
bias is guided by introducing the ORP for the OP. Depending on the
nature of the particular OP, the ORP or the GF must be selected
differently. Besides the cost function or the energy function,
other important properties of the problem is also considered.






Because of the constraint of the ORP or GF, the CS  to be searched
is greatly reduced as the system gets to lower and lower energy.
Thus less computing time is used in our method.


\begin{references}
\bibitem{note1} For a protein chain of length $N$, there are only $N-4$
segments since the first two monomers at each end do not have
segments to be associated with.
\end{references}
\end{document}